# Magic angle effects in a trigonal $Mn_3^{III}$ cluster: deconstruction of a single-molecule magnet

Jonathan Marbey,[1] Pei-Rung Gan,[2] En-Che Yang[2] and Stephen Hill[1,*]

[1]Department of Physics and National High Magnetic Field Laboratory, Florida State University, 1800 East Paul Dirac Drive, Tallahassee, FL 32310, USA

[2]Department of Chemistry, Fu Jen Catholic University, Hsinchuang, New Taipei City, 24205, Taiwan, Republic of China

## ABSTRACT

We present angle-dependent high-frequency EPR studies on a single-crystal of a trigonal $Mn_3^{III}$ cluster with an unusual structure in which the local magnetic easy-axes of the constituent $Mn^{III}$ ions are tilted significantly away from the molecular $C_3$ axis towards the 'magic-angle' of 54.7 degrees, resulting in an almost complete cancelation of the 2$^{nd}$-order axial magnetic anisotropy, $D\hat{S}_z^2$, associated with the ferromagnetically coupled total spin $S_T = 6$ ground state. This contrasts the situation in many related $Mn_3^{III}$ single-molecule magnets (SMMs) that have been studied intensively in the past, for which the local $Mn^{III}$ anisotropy tensors are reasonably parallel, resulting in substantial barriers to magnetization relaxation ($U_{eff}$ ~ 30–35 cm$^{−1}$) and magnetization blocking below about 2.5 K. The suppression of the 2$^{nd}$-order anisotropy [note that the rhombic term, $E(\hat{S}_x^2 - \hat{S}_y^2)$, is also zero on symmetry grounds] in the present case results in a situation in which the zero-field splitting (ZFS) of the $S_T = 6$ ground state is dominated by 4$^{th}$- and higher-order interactions. This provides a unique opportunity to study in depth how molecular geometry

---

influences these interactions that are responsible for quantum tunneling of magnetization in high-symmetry SMMs. Angle-dependent EPR measurements provide a full mapping of the molecular magneto-anisotropy. Meanwhile, irreducible tensor operator (ITO) methods are employed in order to obtain analytic expressions that directly relate molecular anisotropy to the microscopic physics, i.e., the ZFS tensors associated with the individual $Mn^{III}$ ions, their orientations, and the exchange coupling between the three spins. The ITO methodology improves significantly upon previous numerical methods that have been applied to trigonal SMMs. We find that the magic-angle tilting leads to a massive compression of the $S_T = 6$ ground state energy level diagram ($< 3.5$ cm$^{-1}$ separate the lowest and highest lying levels in zero-field) and strong mixing between spin projection states. Although these characteristics are antagonistic to SMM behavior, they provide important insights into the physics of polynuclear molecular nanomagnets.

# I. INTRODUCTION

The synthesis of bistable magnetic molecules, or single-molecule magnets (SMMs), relies on the ability to control the microscopic structural details that dictate the overall molecular magnetic anisotropy.[1-11] This anisotropy lifts the degeneracy of spin states in the absence of an applied magnetic field and, in certain axial geometries, can generate an energy barrier separating spin-up and down states.[1,12] In these cases, magnetic information can effectively be stored in the polarization state of the molecule, provided: (i) the barrier is large compared to $k_B T$; and (ii) quantum tunneling through the barrier can be avoided.[13] Both the barrier height and quantum tunneling are strongly influenced by molecular structure/symmetry, thus motivating detailed studies of structure-property relations.[10]

One approach to creating SMMs involves assembling multiple paramagnetic ions possessing appreciable magnetic anisotropy into larger high-symmetry, high-spin molecules.[1,14-18] The design of these types of molecular spin clusters was historically motivated by the desire of increasing the energy barrier, $U_{\text{eff}} \approx |DS_T^2|$, separating the maximally projected spin-up and down states, where $D$ parameterizes the 2$^{nd}$-order uniaxial anisotropy (through the effective spin Hamiltonian, $\hat{H}_{\text{axial}} = D\hat{S}_z^2$, where $\hat{S}_z$ is the $z$-component spin operator) and $S_T$ the total spin associated with the magnetic ground state of the molecule.[1,12] Naively, increasing $S_T$ may at first sight seem appealing, since the magnitude of the parabolic energy barrier is proportional to $S_T^2$. However, this requires coupling multiple anisotropic ions, which is not only synthetically challenging, but also leads to a dilution of the 2$^{nd}$ order axial molecular anisotropy that, in the best case, scales as $D \propto 1/S_T$.[6,19-23] Moreover, the relatively weak exchange found in most transition metal clusters results in the emergence of higher-order corrections ($\propto \hat{S}_\mu^4$, $\hat{S}_\mu^6$, etc., where $\mu = x, y, z$) to the parabolic 2$^{nd}$ order anisotropy that: (i) ultimately produce completely different energy landscapes; and (ii) may induce new quantum tunneling of magnetization (QTM) pathways.[7-10] These higher order anisotropies arise most prominently when the inter-ion exchange is comparable to (or weaker than) the local single-ion anisotropies; the transverse terms (i.e. $\mu = x, y$) typically also require a tilting of the local anisotropy tensors away from the molecular symmetry ($z$-) axis.[7,8,21,24] In such situations, higher lying spin multiplets effectively admix with the ground spin state, resulting in the aforementioned higher order corrections to the 2$^{nd}$ order anisotropy. However, molecular symmetry still dictates which interactions are allowed. In this regard, trigonal molecules provide a relatively simple case that can effectively demonstrate the interplay between molecular structure, exchange, and local single-ion anisotropy.[7-11] Importantly, QTM is strictly forbidden for a purely 2$^{nd}$ order trigonal spin Hamiltonian, but may become rather strong in the weak exchange limit due to the emergence of symmetry allowed 4$^{th}$ and 6$^{th}$ order transverse interactions.[8] Here, we present a rare example in which the structure of a trigonal $Mn_3$ molecule results in a near total suppression of the 2$^{nd}$ order molecular anisotropy, such that the resultant magnetic and spectroscopic properties are dominated by 4$^{th}$ and higher order interactions.

The article is organized as follows. An overview of the $Mn_3$ molecule is given in Section II, highlighting the important structural features that give rise to the three-fold pattern observed in the electron paramagnetic resonance (EPR) measurements described in Section III. Simulations of the experimental EPR results based on the so-called Giant Spin Approximation (GSA) are presented in Section IV, revealing a significant trigonal contribution to the magnetic anisotropy, and an unusually small 2$^{nd}$-order axial term. These findings require consideration of a Multi-Spin (MS) description, in which the local anisotropies of the constituent atoms are considered along with the exchange coupling between them. This framework is introduced in Section V, along with a mapping between the MS and GSA Hamiltonians using an irreducible tensor operator representation, providing important microscopic insights into the structural factors that influence the total molecular anisotropy. Conclusions are then presented in Section VI.

## II. $Mn_3^{III}$ MOLECULAR STRUCTURE

This investigation focuses on a ferromagnetically coupled $Mn_3^{III}$ triangular molecule (see Fig. 1) displaying rigorous $C_3$ symmetry,[25] in which each $Mn^{III}$ site hosts a local spin $s = 2$, yielding a total molecular ground spin state of $S_T = 6$. The octahedrally coordinated $Mn^{III}$ sites are arranged in such a way that the principal (easy-) axes of the individual anisotropy tensors are tilted significantly (~54°) away from the molecular $C_3$ axis towards the trigonal plane. These axes are defined by the Jahn-Teller (JT) elongated Mn···O bonds lying along the black arrows in Fig. 1(a). Importantly, the JT distortion generates a local 2$^{nd}$ order axial anisotropy of the form $\widehat{H}_{local} = d\hat{s}_{z_i}^2$, where $d$ parameterizes the interaction strength and $z_i$ the orientation of the local axial interaction at each $Mn^{III}$ site, $i$; lowercase symbols are employed here in order to distinguish the parameters from those employed in the molecular Hamiltonian (Section IV). As will be shown, the relatively large easy-axis tilt, which approaches the 'magic angle' of 54.7°, suppresses most of the 2$^{nd}$ order molecular anisotropy, while also giving rise to 4$^{th}$ order trigonal (and 6$^{th}$ order hexagonal) terms that emerge within the coupled molecular spin Hamiltonian. The magic angle, defined as the angle for which the second order Legendre Polynomial, $P_2(\cos\theta) = 0$, holds particular significance in magnetism. As seen in this study, suppression of the 2$^{nd}$-order anisotropy due to the $Mn^{III}$ easy-axis tilting close to the magic angle affords unprecedented sensitivity to higher-order anisotropies via high-field/frequency EPR (HFEPR) measurements.

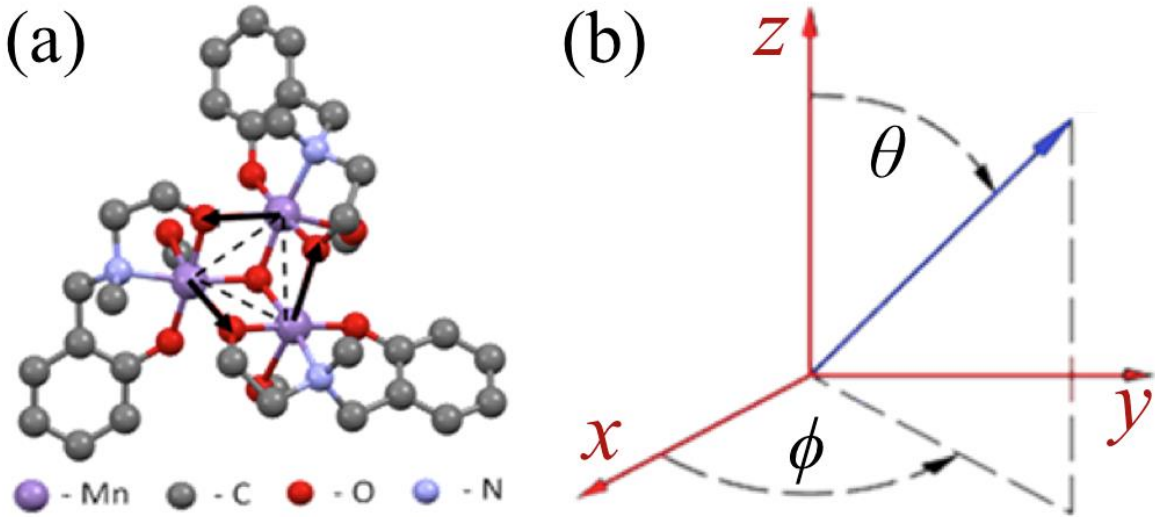

**FIG. 1.** (a) The [Mn$_3$O(mheap)$_3$(CH$_3$OH)$_3$](ClO$_4$) molecule[25] consists of three octahedrally coordinated $Mn^{III}$ ions, each with four unpaired electrons occupying 3$d$ orbitals, yielding a total spin of $s = 2$ at each site. These spins are ferromagnetically coupled via superexchange through the oxygen bridges, giving rise to a giant spin ground state of $S_T = 6$. The black arrows depict the approximate directions of the local easy-axes. (b) Schematic defining the applied magnetic field orientation within the laboratory frame; the single-crystal sample was oriented such that the plane of the Mn$_3$ triangle was approximately in the $xy$ plane of the lab frame; $\theta$ and $\phi$ describe polar and azimuthal field rotation angles.

### III. HIGH FIELD EPR STUDIES

HFEPR measurements were performed using a cavity perturbation technique, with a Millimeter-wave Vector Network Analyzer (MVNA) employed as a source and detector.[26,27] A single-crystal of [Mn$_3$O(mheap)$_3$(CH$_3$OH)$_3$](ClO$_4$) [hereon Mn$_3$, Fig. 1(a)],[25] which crystalizes in the P$\bar{3}$ space group, was mounted in a cylindrical microwave cavity situated within the bore of a 9-5-1 T vector magnet such that the magnetic field direction could be varied in both the polar ($\theta$) and azimuthal ($\phi$) directions with respect to the sample [see Fig. 1(b)].[28] Temperature control was achieved using a variable-flow helium gas cryostat. Details concerning the synthesis, crystal structure and magnetic properties of the Mn$_3$ compound will be published elsewhere.[25] Since Mn$_3$ crystallizes in the form of hexagonal shaped plates, a single-crystal could be mounted such that the trigonal plane formed by the three $Mn^{III}$ ions was approximately co-planar with the $xy$ plane of the lab frame. HFEPR spectra were then recorded at 89.2 GHz, in 10° steps in both $\theta$ and $\phi$, in order to obtain a complete mapping of the molecular anisotropy (see Fig. 1b for definition of coordinates). The stacked plots in Fig. 2(a) and (b), with their accompanying simulations, show example field sweeps for values of $\theta$ spanning a full 180°, and two fixed orientations in $\phi$ [(a) 40° and (b) 100°].

From the stacked plots in Fig. 2, we observe that the position of the strongest EPR transition (assumed to correspond to the excitation from the ground state) does not vary with field orientation as $\sin^2\theta$, as would be expected for a SMM in which the molecular anisotropy is dominated by the 2$^{nd}$-order axial term, $\hat{H}_{\text{axial}} = D\hat{S}_z^2$. Rather, it displays multiple turning points and a clear azimuthal ($\phi$-) dependence: for $\phi = 40°$, the maximum field position peaks at $\theta = 60°$, i.e., 30° above the *xy*- plane; for $\phi = 100°$, the maximum shifts to $\theta = 120°$, i.e., 30° below the *xy*-plane. This already suggests that, compared to Mn$_3$ triangles studied in the past,[2,3,5-10,29] the molecular anisotropy is profoundly influenced by higher-order interactions. In order to investigate this further, the resonance position of the strongest EPR transition was mapped with respect to both $\theta$ and $\phi$, as shown in the color map in Fig. 3(a); additional weaker resonances, which are due to thermal population of excited states within the $S_T = 6$ ground multiplet, were not considered in this figure. The apparent three-fold pattern effectively demonstrates the molecular $C_3$ symmetry inherent to the system [the slow oscillation with 360° periodicity in $\phi$ is due to a few degrees of unavoidable misalignment of the flat crystal with respect to the lab frame; this is addressed in the simulation in Fig. 3(b)]. Rather than having an easy-axis/hard-plane type of anisotropy typical for a SMM, the color map instead reveals multiple hard directions located above and below the trigonal plane, albeit maintaining a $C_3$ symmetry. The microscopic origin of the magnetic anisotropy that gives rise to this behavior will be the main focus of the remainder of this paper.

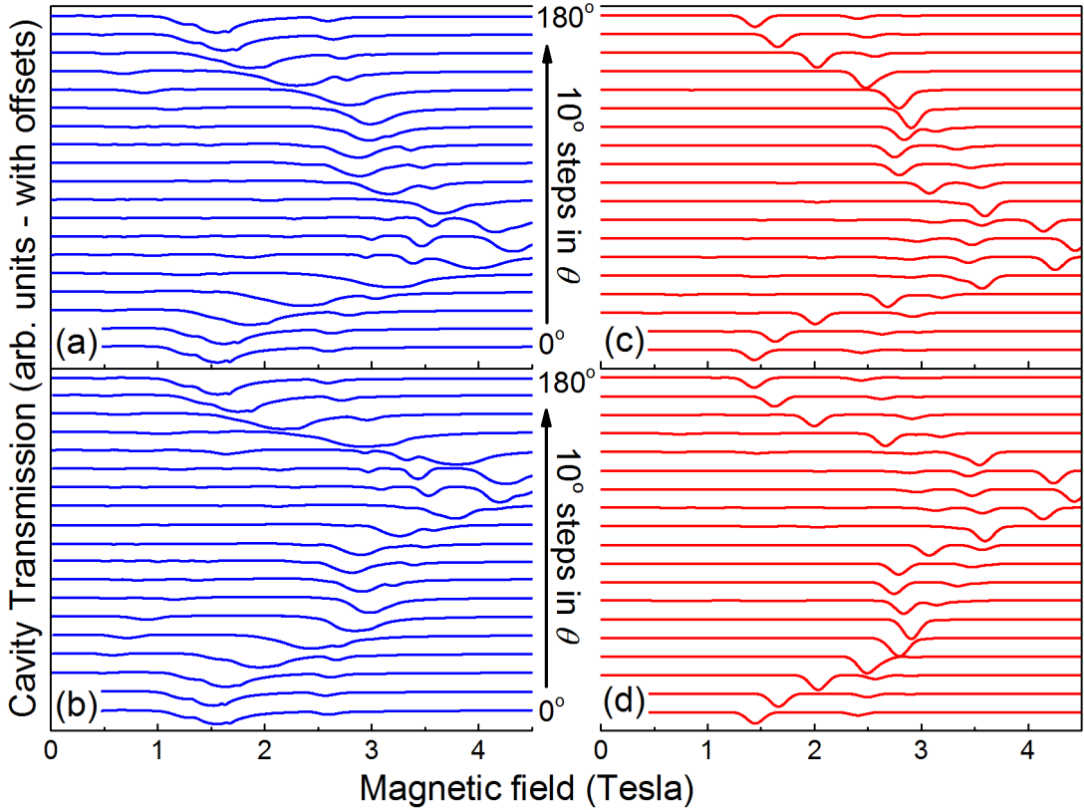

**FIG. 2.** (Left, blue) Experimental spectra collected at 89.2 GHz, in 10° steps of $\theta$, for two azimuthal planes of rotation: (a) $\phi = 40°$, and (b) $\phi = 100°$. Dips in cavity transmission correspond to EPR absorptions. The measurements were performed at 1.65 K (base temperature of the cryostat) in order to minimize thermal population of excited states within the $S_T = 6$ ground multiplet; hence the strongest resonance is assumed to correspond to an excitation from the ground state. (Right, red) Simulations of the experimental spectra at (c) $\phi = 40°$, and (d) $\phi = 100°$, generated using the program EasySpin[30] according to the molecular giant spin Hamiltonian described in the following sections.

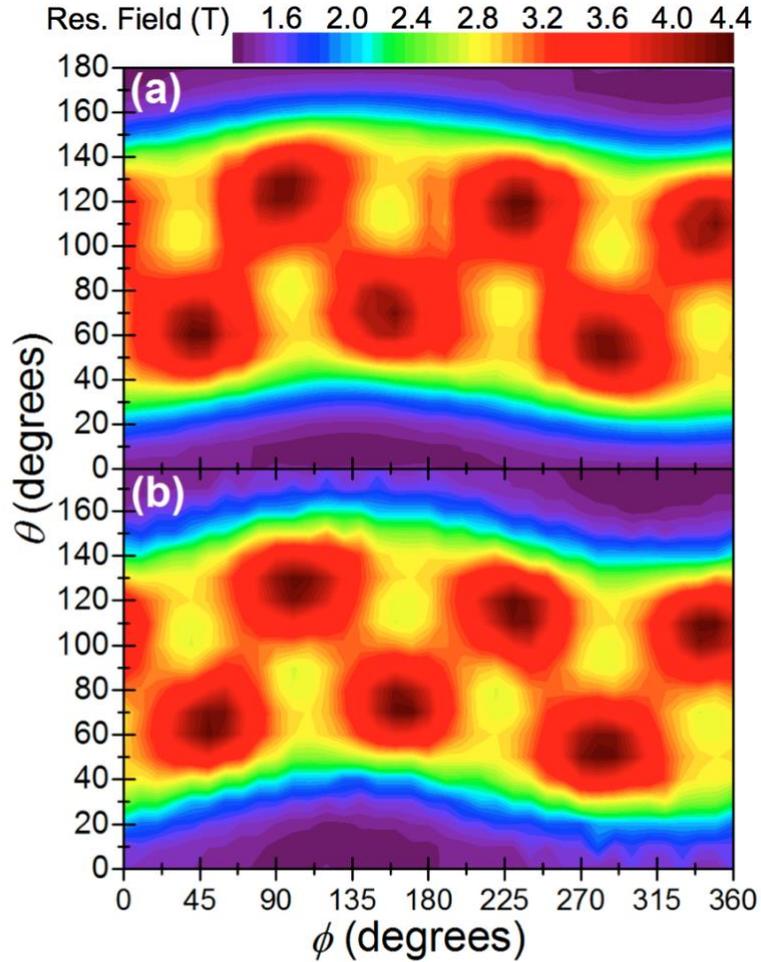

**FIG. 3.** (a) Color map of the position of the strongest EPR transition at 89.2 GHz as a function of $\theta$ and $\phi$; the positions were determined from data sets such as those shown in Fig 2. (b) Simulation of the color map in (a), generated according to the GSA Hamiltonian in Eq. 2, with the parameters listed in Table 1. In order to account for the slow (360° periodicity) oscillation with respect to $\phi$, the simulations were computed on a spherical grid that was iteratively rotated in order to reproduce the small mis-alignment between the crystal and lab coordinate frames; the best simulation was then determined via minimization of the residual with respect to the experimental data. After the minimization, it was found that to match the phase of the trigonal anisotropy terms of the simulation to the lab frame, an additional azimuthal offset of 21.5 ° was required.

## IV. GIANT SPIN MODEL

For the case of exchange-coupled spins, it is common practice to describe a magnetic molecule using an effective Hamiltonian given by the GSA. In the case of Mn$_3$ containing three $s = 2$ sites, this description is particularly advantageous since only the $S_T = 6$ multiplet need be considered, where the lowest $(2S_T + 1) = 13$ energy levels contain the majority of the Boltzmann population at low temperatures; this of course assumes strong ferromagnetic coupling, so that the $S_T = 6$ ground state is well isolated. Such an approach is computationally convenient when compared to the MS description, which requires consideration of $(2s + 1)^n = 125$ states, where n = 3 in the current case.

The Zero-Field Splitting (ZFS) of the ground multiplet is well described by an expansion of the GSA Hamiltonian in terms of Extended Stevens Operators (ESOs):[30,31]

$$\hat{H}_{\mathrm{GSA}} = \sum_{k=2,4,6\ldots} \sum_{q=-k}^{k} B_k^q \, \hat{O}_k^q \tag{1}$$

The $\hat{O}_k^q$ terms are comprised of spin operators of rank $k$, with $q$ specifying the rotational symmetry, which are parameterized by their accompanying $B_k^q$ coefficients. The sum includes non-zero contributions for $k \leq 2S_T$, with the more familiar 2$^{nd}$-order parameters $D = 3B_2^0$ and $E = B_2^2$, where $E$ describes any rhombicity. Note here that the inclusion of higher order ESOs in the GSA gives a strictly phenomenological description of the ZFS. However, this approach has become commonplace in the description of polynuclear clusters,[1,12] for which significant higher order anisotropies often arise due to '$S$-mixing', i.e., admixing of excited $S_T$-multiplets into the ground state due to weak intramolecular exchange coupling. The formalism in Eq. (1) was originally developed to describe the energy levels of isolated magnetic ions in crystalline electric fields as an alternative to lengthy expressions involving linear combinations of tesseral harmonics (for a complete review see M.T. Hutchings).[32] Consequently, as applied in the present context, the terms in Eq. 1 need to reflect the symmetry of the molecule, since the operators themselves contain inherent symmetries. Thus, for a system having rigorous C$_3$ symmetry, one can choose terms containing $\hat{O}_k^{q=3m}$, $m$ being a positive integer. As such, one arrives at the following expression for the molecular ZFS Hamiltonian:[10,31]

$$\hat{H}_{\text{GSA,Mn}_3} = D\hat{S}_z^2 + B_4^0 \hat{O}_4^0 + B_4^3 \hat{O}_4^3 + B_6^3 \hat{O}_6^3 + B_6^6 \hat{O}_6^6 \qquad (2)$$

From the above symmetry related arguments, this expansion could include terms up to $k = 12$. However, including axial terms ($q = 0$) up to $k = 4$ and off-diagonal terms ($q > 0$) up to $k = 6$ provides a satisfactory description of the ZFS for Mn$_3$ (*vide infra*). From the above experimental results, the obtained best simulation parameters are given in Table 1, where we have additionally imposed the condition $q \geq 0$ in order to best match the azimuthal phase of the simulation with respect to the lab frame. Results for three similar C$_3$ symmetric Mn$_3$ triangles are also listed for comparison,[3,8,9] and will be discussed further below.

To have a functional SMM, it is desirable to have a large negative $D$ while minimizing all of the off-diagonal ($q \neq 0$) terms to prevent QTM via mixing of spin projection states, particularly those on opposite sides of the barrier. However, in the present case, the near magic angle tilting of the local Mn$^{III}$ ZFS tensors in fact acts to suppress the 2$^{nd}$-order axial anisotropy, $D$, while giving rise to sizeable 4$^{th}$ and 6$^{th}$ order trigonal (and hexagonal) $q = 3$ ($q = 6$) terms. To understand how this comes about, it is necessary to move to a multi-spin description.

## V. MULTI-SPIN MODEL

For $N$ spin sites, the exchange coupled multi-spin Hamiltonian is given by:[8,10,19,33]

$$\hat{H}_{MS} = \sum_{i=1}^{N} \hat{\mathbf{s}}_i \cdot \overleftrightarrow{R}_i^T \cdot \overleftrightarrow{d} \cdot \overleftrightarrow{R}_i \cdot \hat{\mathbf{s}}_i + \sum_{i,j>i}^{N} J_{ij} \hat{\mathbf{s}}_i \cdot \hat{\mathbf{s}}_j; \quad \overleftrightarrow{d} = \begin{pmatrix} -\dfrac{d}{3} + e & 0 & 0 \\ 0 & -\dfrac{d}{3} - e & 0 \\ 0 & 0 & \dfrac{2d}{3} \end{pmatrix}, \quad (3)$$

where the indices $i$ and $j$ refer to the $N$ sites in the molecule, with associated spin $\hat{s}_i$, and local 2$^{nd}$-order anisotropy specified by the tensor $\overleftrightarrow{d}$ (assumed to be the same at all three sites in the present case due to the C$_3$ symmetry); $d$ and $e$ are respectively the local 2$^{nd}$-order axial and rhombic anisotropy parameters. The $\overleftrightarrow{R}_i$ tensors represent Euler rotation matrices, specified by angles $\alpha_i$, $\beta_i$, and $\gamma_i$, that relate the local coordinate frame of each ion to the molecular (lab) frame. Specifying the Euler angles thus allows us to impose symmetry constraints on the system; note here that we use the '*zyz*' Euler convention.[30]

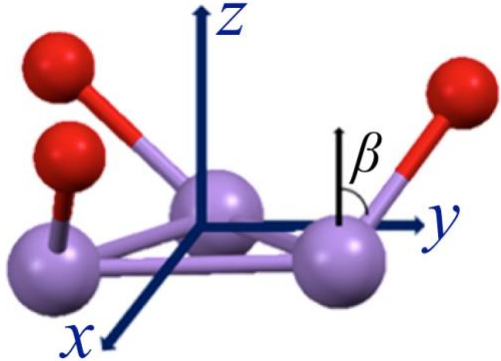

**FIG. 4.** Simplified model of the Mn$_3$ molecule. The $\overleftrightarrow{d}$ tensor of each Mn$^{III}$ ion is related to its neighbor by a rotation of $\alpha = 120°$, with $\beta$ specifying the tilt from the molecular C$_3$ ($z$-) axis.

In order to demonstrate how molecular symmetry affects the total magnetic anisotropy of the Mn$_3$ system, we model it as an equilateral triangle of ferromagnetically coupled $\hat{s}_i = 2$ spins with axial $\overleftrightarrow{d}$ tensors (initially keeping $e = 0$ for the sake of simplicity), fixing $\alpha_1 = 0$, $\alpha_2 = 120°$, $\alpha_3 = 240°$, and further imposing $\beta_1 = \beta_2 = \beta_3 = \beta$ and $\gamma_1 = \gamma_2 = \gamma_3 = 0$ to preserve the C$_3$ symmetry. In general, the spin Hamiltonian has a symmetry that is higher than the spatial symmetry of the molecule. For example, the special case of $\beta = 0$ gives a MS Hamiltonian with cylindrical D$_{\infty h}$ symmetry, because all of the local anisotropy projects onto a single axis (the molecular C$_3$ axis). For $\beta = 90°$, the MS Hamiltonian has a hexagonal D$_{6h}$ symmetry while, for all $0 < \beta < 90°$, the MS Hamiltonian adopts an D$_{3d}$ symmetry. These latter two cases acquire a higher symmetry than the molecule because of the additional time-reversal invariance of the spin-orbit interaction.[10] Inclusion of a finite *e* parameter reduces the symmetry in some cases, e.g., $\beta = 0$ reduces to D$_{6h}$ symmetry. However, in the absence of an applied field, the MS Hamiltonian never reduces to the C$_3$ symmetry of the molecule. Based on these simple arguments, one can immediately predict which terms in the GSA Hamiltonian to expect in various limiting cases. For example, the $q = 0$ ESOs all have D$_{\infty h}$ symmetry, the $\hat{O}_6^6$ operator possesses a hexagonal D$_{6h}$ symmetry, while the remaining $\hat{O}_4^3$ and $\hat{O}_6^3$ operators both impose an D$_{3d}$ symmetry on the GSA Hamiltonian.

While the above qualitative arguments are appealing, we seek a quantitative correspondence between the MS and GSA parameterizations. To accomplish this, we take a perturbative approach following the procedure developed by Waldmann and Güdel,[34] expressing the total molecular ZFS in terms of its equivalent multi-spin operators. This permits investigation of how both the orientations and magnitudes of the local anisotropies influence the molecular anisotropy. Although not as precise as performing exact matrix diagonalizations, as in the case of Liu et al.,[8] this approach allows us to derive analytical expressions connecting the *microscopic* MS and *effective* GSA models.

We first reframe the problem by casting the local anisotropies as a perturbation to the isotropic exchange coupling. This allows us to calculate the matrix elements from the MS Hamiltonian up to 2nd order, such that equivalent operators can be generated in the subspace where:

$$\langle SM|\hat{H}_{GSA}|SM'\rangle = \langle \tau SM|\hat{H}_{MS}|\tau'S'M'\rangle, \tag{4}$$

in which $\tau S$ specifies a single spin multiplet within the full Hilbert space described by the multi-spin basis, while $|SM\rangle$ specifies a single spin subspace spanned by the GSA introduced in Eq. (2). For a given $S$ in the giant spin subspace, the case $M = M'$ describes diagonal matrix elements. Similarly, matrix elements originating from within the same spin-multiplet have the same $\tau$ and $S$. Here, the label $\tau$ completely specifies the spin state in the MS description, and serves to simplify the more conventional coupled basis given by $|S_1 S_2 S_{12} S_3 SM\rangle$.[19,33] The procedure for finding equivalent operators thus involves taking projections of the following form for each term in the expansion:

$$\hat{H}_{GSA} = P_{\tau S}\hat{H}_{MS}P_{\tau'S'} = \sum_{M,M'}|SM\rangle\langle \tau SM|\hat{H}_{MS}|\tau'S'M'\rangle\langle SM'|$$

where,

$$P_{\tau S} = \sum_{M}|SM\rangle\langle \tau SM|$$

(5)

in which the sum is taken over the GSA substates. We begin by breaking up the MS Hamiltonian as:

$$\hat{H}_{MS} = \hat{H}_0 + \hat{H}_1, \tag{6}$$

where the $0^{\text{th}}$ order isotropic exchange interaction is given by $\hat{H}_0$ [$2^{\text{nd}}$ term in Eq. (3)], and $\hat{H}_1$ specifies the local anisotropies parameterized by $d$ and $e$ [$1^{\text{st}}$ term in Eq. (3)]. From here, the matrix elements can be written according to the following expansion:[35,36]

$$\langle \tau S M | \hat{H}_{\text{MS}} | \tau S M' \rangle$$
$$= E_{0\tau S} + \langle \tau S M | H_1 | \tau S M' \rangle$$
$$+ \sum_{\tau'', S'', M''} \frac{\langle \tau S M | H_1 | \tau'' S'' M'' \rangle \langle \tau'' S'' M'' | H_1 | \tau S M' \rangle}{E_{0\tau'' S''} - E_{0\tau S}} \quad (7)$$

in which the leading term gives the energy due to exchange ($\hat{H}_0$), the $1^{\text{st}}$ order perturbation considers mixing of $M$ states due to *intra*-spin multiplet matrix elements (for a given $\tau S$ state), while the $2^{\text{nd}}$ order perturbation considers *inter*-multiplet mixing between states $\tau S$ and $\tau'' S''$, as illustrated in the Zeeman diagram of Fig. 5 (note that the Zeeman interaction is not explicitly included in the above zero-field expressions). *Inter*-multiplet mixing, more commonly referred to as '*S*-mixing', is often considered to be weak in comparison to $1^{\text{st}}$ order mixing in molecular spin systems. Such assumptions are based on the notion of well-separated spin states [large denominator in the $2^{\text{nd}}$ order perturbation in Eq. (7)], particularly in ferromagnetic cases with appreciable exchange ($J > d$). However, as will be shown in the following sections, the suppression of $1^{\text{st}}$ order mixing due to the magic-angle tilting of the local $\overleftrightarrow{d}$ tensors in the Mn$_3$ molecule considered here requires careful consideration of the $2^{\text{nd}}$-order terms. Those not interested in the explicit derivations, starting from Eq. (7), may proceed directly to the results given by Eqs. (8) and (9), which respectively consider the $k = 2$ and $k = 4$ anisotropy terms ($1^{\text{st}}$ and $2^{\text{nd}}$-order perturbations) in the GSA Hamiltonian. The derivations involving irreducible tensor operator methods are addressed in Appendix A.

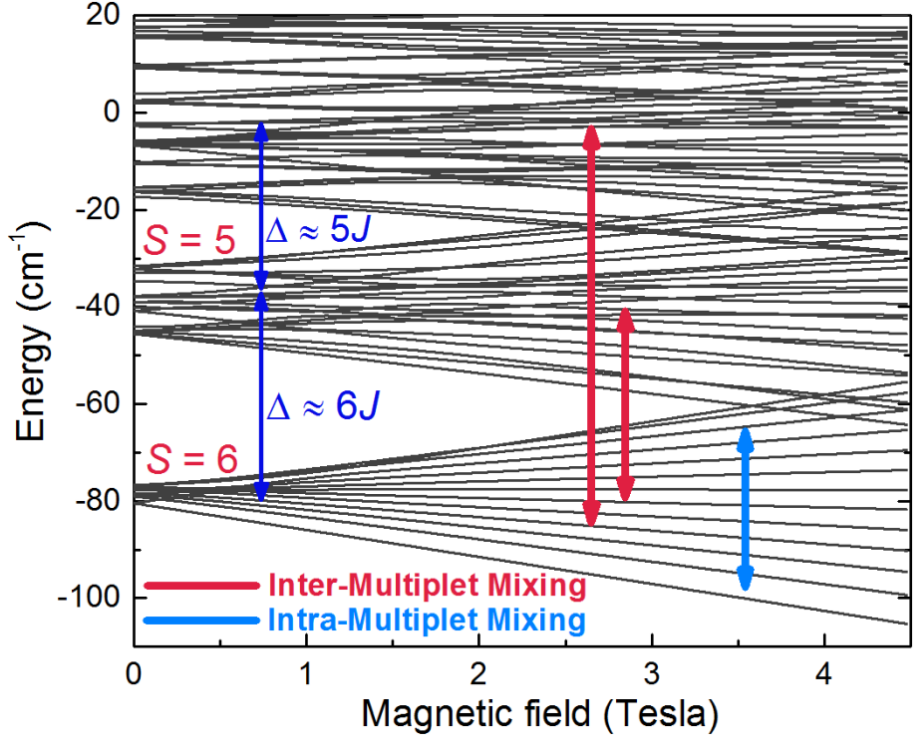

**FIG. 5.** Zeeman energy level diagram for the lowest lying spin multiplets, with $\beta$ close to the magic angle. According to the perturbative approach in Eq. (7), the $0^{th}$ order energy splitting ($\Delta$) between spin multiplets is due purely to the exchange Hamiltonian. Meanwhile, the $1^{st}$ and $2^{nd}$ order ZFS interactions are due to intra- and inter-Multiplet spin-state mixing, respectively (see legend). Note that, while there exist many higher lying excited spin multiplets, the second order treatment discussed here mixes only the doubly degenerate $S = 5$ and triply degenerate $S = 4$ multiplets into the $S = 6$ ground spin multiplet.

### A. First order perturbation

In terms of the MS anisotropy parameterization in Eq. (3), consideration of *intra*-multiplet mixing yields the following expression for the $k = 2$ contribution to the GSA anisotropy:

$$D_{\text{mol}}(\beta) = 3B_2^0 = \frac{3d}{2}\Gamma_{k=2}(s)(3\cos^2\beta - 1), \quad (8)$$

where the projection factor described in Eq. (A9) has $\Gamma_{k=2}(s) \approx .0909$ for $s = 2$ and $S = 6$ (see expressions in Table 11.9 of Boča).[33] Note here that, for the sake of simplicity, we have neglected any contribution from local rhombic distortions in the perturbative calculation, i.e., we set $e = 0$. Moreover, the molecular C$_3$ symmetry dictates that the $k = 2$, $q = 2$ GSA parameter $E = B_2^2 = 0$.

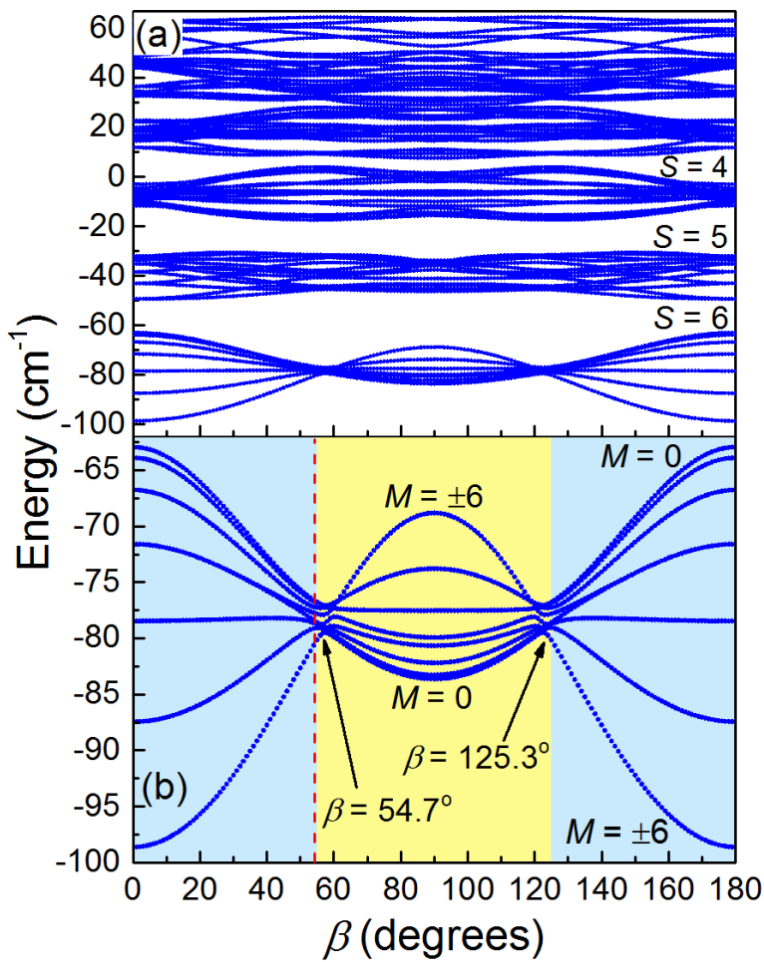

**FIG. 6.** (a) Energy eigenvalues obtained via exact diagonalization of Eq. (3) using the MS parameters $J = -6.35$ cm$^{-1}$ $d = -3.73$ cm$^{-1}$ and $e = 0$, plotted as a function of the tilt angle $\beta$. (b) Expanded view of the ground $S_T = 6$ spin multiplet; note that the eigenvalues converge at the two magic angles, $\beta = 54.7°$ and 125.3°, which delineate regions of negative (blue) and positive (yellow) $D_{mol}$. The red dashed line indicates the value of $\beta$ that provides the best overall agreement between experiment and the simulations for this molecule (*vide infra*). Note that, although the ideal axial case is plotted above, the addition of a local rhombic ZFS interaction (finite $e$) will slightly shift where the energy levels converge in $\beta$.

From Eq. (8), one sees that there are two 'magic' angles where $D_{mol}(\beta) = 0$, i.e., $\beta = 54.7°$ and 125.3°. To better understand this, we perform an exact diagonalization of the MS Hamiltonian of Eq. (3) at zero applied magnetic field to generate the energy eigenvalues as a function of $\beta$ (see Fig. 6). Considering only the ground $S_T = 6$ multiplet [Fig. 6(b)], one sees that the eigenvalues converge and become nearly degenerate at the magic angles. Importantly, these angles delineate the boundaries separating regions with opposite signs of the molecular 2$^{nd}$-order ($k = 2$) GSA anisotropy (2$^{nd}$-order refers here to the order of the GSA spin operators, as opposed to the perturbation order): assuming local easy-axis anisotropy ($d < 0$), the single-ion $\overleftrightarrow{d}$ tensors project a net easy-axis anisotropy onto the molecular C$_3$ axis for the regions $\beta < 54.7°$ and $\beta > 125.3°$, giving $D_{mol} < 0$ and $M = \pm 6$ ground states; meanwhile, they project a net easy-plane anisotropy into the molecular *xy*-plane for $54.7° < \beta < 125.3°$, giving $D_{mol} > 0$ and a non-magnetic $M = 0$ ground state. As such, one sees that the tilt angle plays a significant role in determining the total molecular anisotropy of the system, i.e., a tuning of this single parameter can result in entirely different magnetic behavior.

At the magic angle, $\beta = 54.7°$, the 2$^{nd}$-order axial molecular anisotropy is completely suppressed, i.e., $D_{mol} = 0$. However, in spite of this suppression, the energy eigenvalues obtained via exact diagonalization of the MS Hamiltonian avoid complete convergence, as seen in Fig. 6(b). Since $e = 0$, and the C$_3$ molecular symmetry forbids rhombic anisotropy (i.e., $B_2^2 = 0$), this suggests the importance of higher order *inter*-multiplet '*S*-mixing' effects. As noted above, such anisotropies are normally obscured in EPR experiments due to the dominant $k = 2$ contributions to the GSA anisotropy. However, the suppression of $D_{mol}$ in the present case affords a rare opportunity to characterize the higher order trigonal anisotropy via EPR, thus justifying the need to consider the 2$^{nd}$-order (of rank $k = 4$) perturbative expansion of Eq. (7).

### B. Second-order perturbation

We next consider how inter-multiplet *S*-mixing gives rise to $k = 4$ ESOs in the GSA of Eq. (2). In order to generate the molecular ZFS parameters pertaining to Mn$_3$, we focus on the ground $S = 6$ multiplet, and consider mixing contributions from the doubly degenerate $S'' = 5$ and triply degenerate $S'' = 4$ multiplets, which respectively lie $\Delta = 6J$ and $\Delta = 11J$ above the unperturbed ground state (see Fig. 5); the double prime used here refers to the excited states that mix with the ground spin state. Higher-lying ($S'' < 4$) multiplets do not mix at 2$^{nd}$-order of perturbation. Our goal here is to determine how the tilt angle $\beta$ influences the strength of the trigonal $\hat{O}_4^3$ interaction. The 2$^{nd}$ order perturbation derivation addressed in Appendix A gives:

$$B_4^3(\beta) = \sum_{\tau''S''} \Gamma_{2,4}(s)\Gamma_{2,4}^*(s) d_{1,2,3}^{2,2,4} \frac{3d^2}{\Delta} \cos\beta \sin^3\beta, \tag{9}$$

where the $\Gamma_{k_1,k}$ projection coefficients relate the MS operators of order $k_1 = 2$ to the GSA operators of order $k = 4$; the summation is over excited states, $\tau''S''$, as seen in Eq. (A11) [see also Eq. A.19].

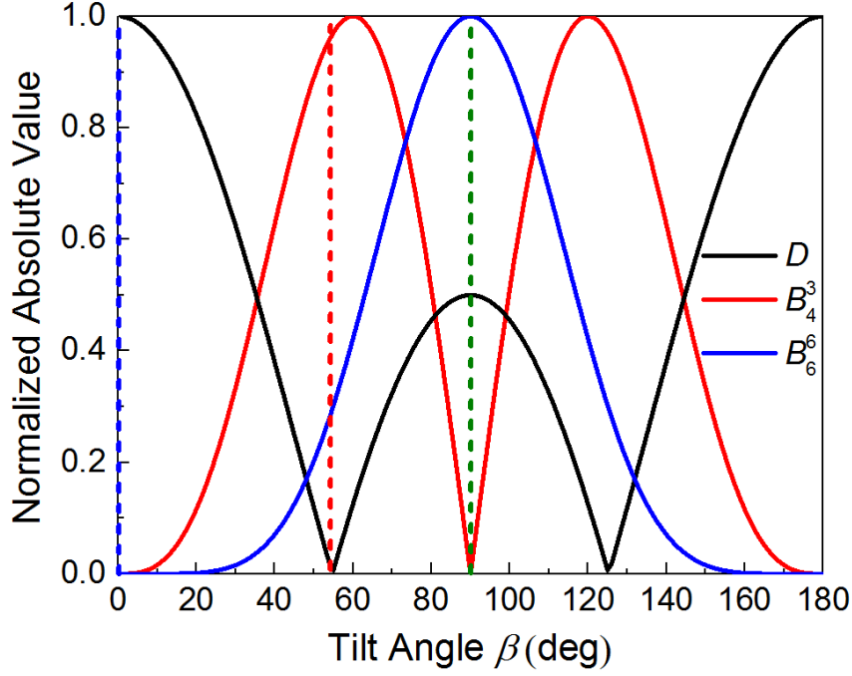

**FIG. 7.** Dependence of the absolute values of the GSA anisotropy parameters $D$, $B_4^3$ and $B_6^6$ on the easy-axis tilt angle, $\beta$, associated with the local Mn$^{III}$ $\overleftrightarrow{d}$-tensors (see Fig. 4). The vertical dashed lines refer to the simulations in Fig. 8, and the curves have been normalized to the maximum values of the parameters. At the tilt angle corresponding to the best simulation ($\beta = 54.3$, red dashed line), the usually dominant $D\hat{S}_z^2$ interaction is almost completely suppressed. Consequently, the 4$^{th}$ order $B_4^3 \hat{O}_4^3$ interaction (2$^{nd}$ order in the perturbation of Eq. 7) makes a comparable contribution to the overall ZFS within the ground $S_T = 6$ spin state, while the $B_6^6 \hat{O}_6^6$ contribution (3$^{rd}$ order perturbation) is about an order of magnitude smaller.

Referencing the tabulated products of Wigner-3j and 6j symbols, given by the $d_{q_1 q_2 q}^{k_1 k_2 k}$ values,[34] and calculating the $\Gamma_{2,4}(s)$ coefficients from the expressions given in Boča,[33] a direct mapping can be made from the MS to the GSA Hamiltonian. It is important to note the inverse dependence on $\Delta$ and, therefore, $J$. Consequently, $B_4^3$ provides a direct route to determining $J$, provided $d$ and $\beta$ are independently known.

Before explicitly relating the experimental GSA parameters in Eq. (2) to the MS parameters in Eq. (3) from the above derivations, we examine the dependence of $D_{\text{mol}}$ and $B_4^3$ on the Mn$^{\text{III}}$ $\overleftrightarrow{d}$ tensor tilt angle $\beta$ for the three cases highlighted by dashed vertical lines in Fig. 7. For completeness, we also consider the contribution from the $B_6^6 \hat{O}_6^6$ interaction, which varies as $d\sin^6(\beta)$ (assuming $e = 0$), and is derived following the previously described steps to 3$^{\text{rd}}$ order in the perturbation.[37] For the MS simulation shown in Fig. 8(a) with $\beta = 0°$, for which all three $\overleftrightarrow{d}$ tensors are aligned with the molecular C$_3$ axis, $B_4^3$ is symmetry forbidden (as noted above), while $D_{\text{mol}}$ is negative and its magnitude is maximum (see Fig. 7). If we also neglect any local rhombicity ($e = 0$), the GSA Hamiltonian acquires a uniaxial D$_{\infty h}$ symmetry that is rotationally invariant with respect to $\phi$ [Fig. 8(a)]. This gives rise to the familiar easy-axis/hard-plane anisotropy that is characteristic of most SMMs. Tilting to the opposite extreme of $\beta = 90°$ yields the opposite result, with an easy-plane/hard-axis type anisotropy and a positive $D_{\text{mol}}$ that is locally maximum at $\beta = 90°$ [with $D_{\text{mol}}(90°) = -\frac{1}{2}D_{\text{mol}}(0°)$]. Note that $B_6^6$ is also maximum for $\beta = 90°$, which produces a weak 6-fold modulation of the anisotropy in the hard-plane. This rank-6 term is responsible for the D$_{6h}$ symmetry reflected in the simulated color map [Fig. 8(c)]. Like the $\beta = 0°$ case, $B_4^3$ is symmetry forbidden for $\beta = 90°$ (provided $e = 0$). However, tilting away from these extremes results in the emergence of the $B_4^3 \hat{O}_4^3$ interaction, which attains its maximum at $\beta = 60°$, i.e., close to the magic angle (54.7°) at which $D_{mol}$ is exactly zero. This coincidence leads to a situation in which the simulations are extremely sensitive to $\beta$ in the vicinity of the magic angle, because it strongly influences the relative mixture of interactions that produce axial ($D_{\infty h}$ symmetry) and trigonal ($D_{3d}$ symmetry) modulations of the EPR peak positions. This sensitivity provides tight constraints on the optimal value of $\beta = 54.3°$ [Fig. 8(b)] employed in the best MS model simulation (*vide infra*).

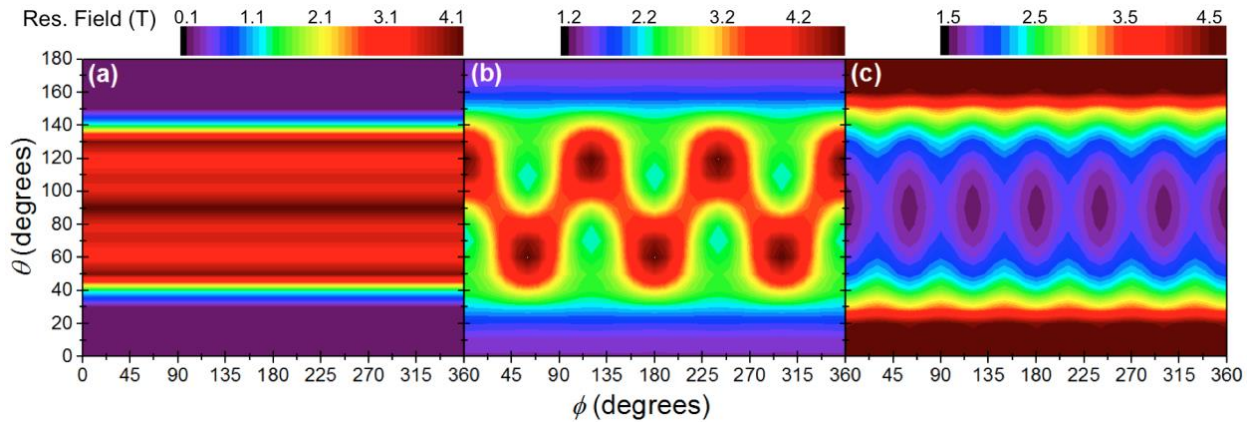

**FIG. 8.** MS simulated color maps for the ground state resonance position for the three different easy-axis tilts highlighted in Fig. 7: (a) $\beta = 0°$, (b) $\beta = 54.3°$, and (c) $\beta = 90°$. Note here that the color scales differ for each map; the intent is to qualitatively highlight the symmetry pattern for each case (see also Fig. 3).

Mapping the experimentally obtained GSA parameters onto the MS model now becomes a matter of choosing appropriate local anisotropy parameters $d$ and $e$, then using the algebraic relations found for $D_{mol}(\beta)$ to constrain the tilt angle $\beta$, and the experimentally obtained value of $B_4^3$ to constrain $J$. Simulations obtained via exact diagonalization of the full $125 \times 125$ MS Hamiltonian give good agreement with the GSA model (and, hence, the experimental data) with local parameters: $d = -3.73$ cm$^{-1}$ and $e = -0.22$ cm$^{-1}$, together with $\beta = 54.3°$ and $J = -6.35$ cm$^{-1}$.

Although the preceding discussion considered the simple case where $e = 0$, the addition of a small rhombic anisotropy is required to improve the overall mapping between the two models. The full expressions that take this additional anisotropy into account are given in Eqs. (A14) and (A19).

The obtained single-ion anisotropy parameters, *d* and *e* (Table 1), are consistent with those deduced from studies of mononuclear Mn$^{III}$ complexes in similar Jahn-Teller distorted octahedral coordination environments,[38] as well as those found for related trinuclear Mn$^{III}$-oxo clusters.[3,6,8,10] Meanwhile, simulations of temperature-dependent magnetic susceptibility data for the present Mn$_3$ compound give an exchange coupling $J = -7.44$ cm$^{-1}$, with *ab initio* calculations suggesting $J = -8.12$ cm$^{-1}$.[25] Moreover, the same *ab initio* calculations suggest that the tilt angle $\beta \sim 60°$. These findings are in good agreement with the present investigation. In fact, examination of the structure in Fig. 1 reveals that the Jahn-Teller elongated O—Mn—O bonds lie along three orthogonal edges of one half of a cube-like structure. Of course, the C$_3$ axis of a perfect cube (the diagonal between opposite corners) is oriented exactly at the magic angle relative to its edges.

When considering previous work on related trigonal Mn$_3^{III}$ molecules,[8,9] we find several striking differences with the present example. For systems with minimal $\overleftrightarrow{d}$ tensor tilting ($\beta < 10°$), the magnitude of $B_4^3$ is substantially smaller than found here. Liu et al.[8] compared examples with identical single-ion ZFS parameters, one with $\beta = 0$ (Triangle 1) and another with $\beta = 8.5°$ (Triangle 2). In the $\beta = 0$ case, it was shown numerically and argued on group theoretic/symmetry grounds that any trigonal GSA terms must be identically zero. On the other hand, for the $\beta = 8.5°$ case, inclusion of a small $B_4^3$ was required to replicate QTM rates measured at certain avoided level crossings.[7] This was also shown to be the case in a more recent investigation by Atkinson et al.,[9] with $\beta = 6°$ (Triangle 3), in which a three-fold pattern of QTM rates could only be explained via inclusion of a $B_4^3 \hat{O}_4^3$ interaction in the GSA Hamiltonian. In the present (more tilted) case, however, the experimentally obtained $B_4^3$ parameter is larger by an order of magnitude compared to the previous examples, while *D* is smaller by an order of magnitude (see Table 1). This is attributed to the tilt of the $\overleftrightarrow{d}$ tensor, where $\beta = 54.3°$ is just 0.4 degrees away from the zero of $D_{\text{mol}}(\beta)$, and only 5.7 degrees away from (or 4% below) the maximum of $B_3^4(\beta)$. Hence, the magnetization reversal barrier is almost completely suppressed, and effects due to *S*-mixing (i.e. QTM) are very pronounced (see below). It is therefore no surprise that the present Mn$_3$ compound does not show any evidence for slow magnetization relaxation at low temperatures.[25]

Similar effects due to tilting of local anisotropy tensors have been discussed in the context of other trigonal magnetic molecules. For example, an Fe$_3$Cr propeller-type complex, for which the single-ion Fe$^{III}$ sites are easy-plane (positive *d*), has been studied extensively by Sorace et al.[11] Here, the individual hard-axes are oriented within 5° of the plane of the Fe$_3$ triangle, thereby projecting an overall easy-axis (negative *D*) anisotropy for the coupled molecule (together with sizeable trigonal terms) – a situation corresponding to an energy inversion of Fig. 6. However, the hard-axis tilting is far from the magic angle, and the magnitude of *D* remains sizeable.

**TABLE 1.** Comparison between the GSA and MS parameterizations of the present tilted Mn$_3$ complex with three closely related molecules (Triangle #1,[8] Triangle #2, [3,8] and Triangle #3[9,39]).

| GSA Parameter | Tilted Mn$_3$ | Triangle #1 | Triangle #2 | Triangle #3 |
|---|---|---|---|---|
| $D$ (cm$^{-1}$) | −0.07 | −0.762 | −0.804 | −0.602 |
| $B_4^0$ (cm$^{-1}$) | $-1 \times 10^{-4}$ | $-1.52 \times 10^{-5}$ | $-5.28 \times 10^{-5}$ | $-2.78 \times 10^{-5}$ |
| $B_4^3$ (cm$^{-1}$) | $-1.09 \times 10^{-3}$ | 0 | $3.32 \times 10^{-4}$ | $-1.99 \times 10^{-4}$ |
| $B_6^3$ (cm$^{-1}$) | $3.34 \times 10^{-6}$ | 0 | -- | 0 |
| $B_6^6$ (cm$^{-1}$) | $1.67 \times 10^{-6}$ | $3 \times 10^{-7}$ | -- | $7.97 \times 10^{-7}$ |

| MS Parameter | Tilted Mn$_3$ | Triangle #1 | Triangle #2 | Triangle #3 |
|---|---|---|---|---|
| $d$ (cm$^{-1}$) | −3.73 | −2.92 | −2.92 | −2.5 |
| $e$ (cm$^{-1}$) | −0.22 | 0.626 | 0.626 | 0.43 |
| $J$ (cm$^{-1}$) | −6.35 | −6.95 | −6.95 | −2.15 |
| $\beta$ (deg) | 54.3 | 0 | 8.5 | 6.0 |

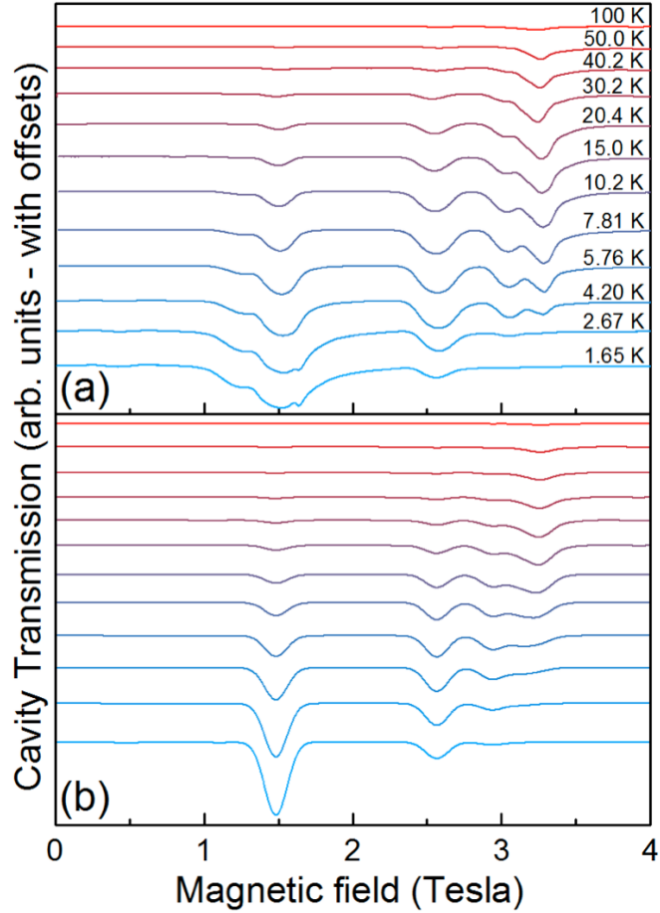

**FIG. 9.** (a) Experimental and (b) simulated 89.2 GHz EPR spectra at $(\theta, \phi) = (0°, 0°)$ as a function of temperature (from 1.65 K to 100 K). The simulations were performed using the MS model and parameters described above.

For completeness, Fig. 9 compares experimental EPR spectra with simulations (for $B//C_3$-axis) that have been generated using the spin-Hamiltonian parameters deduced on the basis of the best simulation of the color map in Fig. 3. These temperature-dependent spectra include transitions between excited spin projection states within the ground $S_T = 6$ multiplet. As such, they provide a far more stringent test of the parameterization because the color maps in Fig. 3 consider only a single EPR transition from the lowest-lying $M$ substate. The broadening and modulation of the lowest field (ground state) resonance in Fig. 9(a) is attributed to intermolecular interactions (the nearest-neighbor Mn-Mn distance is ~10 Å); these interactions are notoriously difficult to simulate, requiring very significant computational resources. The effect is most pronounced at low temperatures due to exchange averaging at elevated temperatures. For this reason, it is only the ground state resonance at ~1.5 T that is significantly affected by intermolecular interactions for temperatures below ~15 K. Overall, the correspondence between experiment and simulations is highly satisfactory, both in terms of the resonance positions and spectral weight (integrated intensity).

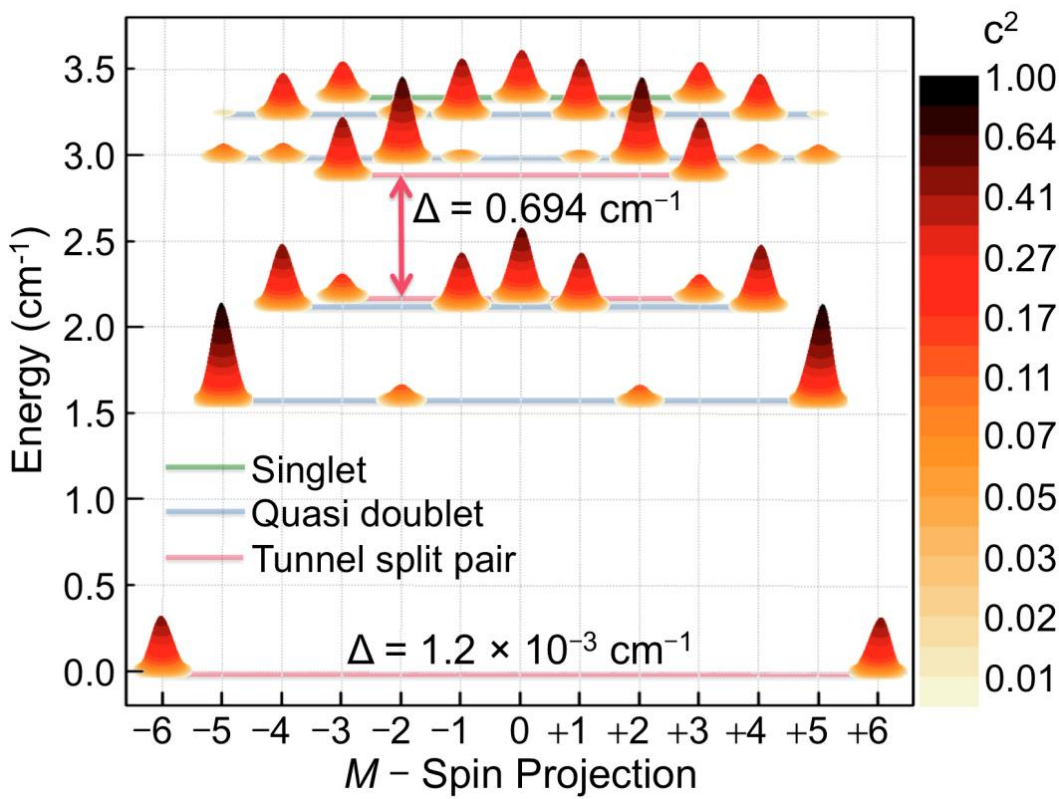

**FIG. 10.** Zero-field energies (referenced to the ground state) and spin projection ($M$) compositions of the 13 eigenstates associated with the $S_T = 6$ ground state of the tilted $Mn_3$ molecule. The color scale and vertical heights of the pillars denote the probability ($c^2$) distributions associated with the eigenstates, which are highly mixed due to the off-diagonal terms in the GSA Hamiltonian (primarily $\hat{O}_4^3$, $\hat{O}_6^3$, $\hat{O}_6^6$) whose anti-commutators contain $(\hat{S}_+^3 + \hat{S}_-^3)$ and $(\hat{S}_+^6 + \hat{S}_-^6)$. The high symmetry of the molecule results in four quasi-doublets (degenerate pairs), a singlet, and two tunnel split doublets. This is the reason for the smaller amplitudes of the ground state pair, as they each consist of near 50:50 ($c^2 = 0.5$) mixtures of the $M = \pm 6$ projections (one symmetric combination, the other antisymmetric). By contrast, the lowest excited quasi doublet consists of one state that is mostly $M = -5$ (with a small admixture of $-2$), and another state that is mostly $M = +5$ (with a small admixture of $+2$). Components that make up less than 1% of the total composition have been omitted for clarity. For example, the ground states have 0.0196% contributions from $M = \pm 3$ and 0.00348% from $M = 0$.

Finally, Fig. 10 plots the zero applied field energies and compositions of the $2S + 1 = 13$ eigenstates associated with the $S_T = 6$ ground state, deduced on the basis of the GSA Hamiltonian, with the parameters given in Table 1. The first obvious thing to note is the very small energy scale of $< 3.5$ cm$^{-1}$ separating the (mostly) $M = \pm 6$ ground states from the highest lying eigenstate. If one naively associates this to a barrier against magnetization relaxation, $U_{eff}$, it is about an order of magnitude smaller than the effective barrier associated with non-tilted trigonal Mn$_3$ SMMs ($U_{eff} \sim 32$ cm$^{-1}$).[3] However, it is also apparent that the eigenstates are strongly mixed in the present case. For example, the tunnel splitting (or QTM gap) associated with the $M = \pm 6$ ground doublet is $1.2 \times 10^{-3}$ cm$^{-1}$, which corresponds to a QTM rate of about 36 MHz, compared to about 400 kHz for the non-tilted case,[7] i.e., a two orders of magnitude difference. Meanwhile, the next tunnel-split pair has an associated gap of 0.69 cm$^{-1}$ (or a tunneling rate of ~21 GHz). Though these tilting effects have a profound influence on the magnitude of the tunneling gaps, the strength of the exchange between neighboring ions is also extremely important in determining the transverse anisotropy. Closer inspection of the tunneling gaps provides an opportunity to test the agreement between the GSA and MS models. Diagonalization of Eq. (3) using the MS parameters in Table 1 gives zero-field tunneling gaps of $7.7 \times 10^{-3}$ cm$^{-1}$ and 0.689 cm$^{-1}$ within the $M = \pm 6$ and $\pm 3$ quasi-doublets, respectively. These values are in fair agreement with those given above on the basis of the GSA model. The discrepancies most likely arise due to our neglect of the $k = 6$ terms in the mapping procedure described in Section V; consideration of $B_6^3$ and $B_6^6$ would require expanding Eq. (7) to 3$^{rd}$ order in the perturbation. Improvement in the mapping in such a case has been extensively demonstrated in analysis of the tunneling gaps in Fe$_3$Cr.[37] The crucial point here is that the large tunneling rates demonstrate that the effective barrier to magnetization relaxation is essentially non-existent for this Mn3 molecule. Meanwhile, we note that large tunneling gaps have recently been shown to be important in the context of quantum technologies.[40]

## VI. CONCLUSIONS

We present the results of detailed two-axis, angle-dependent high-field EPR studies of a single crystal of an unusual $Mn_3^{III}$ triangular nanomagnet displaying rigorous $C_3$ symmetry. Unlike similar triangles studied in the past, the easy-axes of the individual $Mn^{III}$ ions in this particular molecule are tilted very close to the so-called 'magic angle' of $\theta_m = 54.7°$. This combined with the trigonal symmetry results in a situation in which the 1$^{st}$ order spin-orbit anisotropy (quadratic in terms of spin operators, i.e., $\hat{S} \cdot \overleftrightarrow{D} \cdot \hat{S}$) is almost completely suppressed. Consequently, the overall magneto-anisotropy is dominated by 2$^{nd}$ and higher-order trigonal spin-orbit interaction terms (4$^{th}$ and higher order in terms of spin operators). The angle-dependent EPR studies offer a powerful and direct means of visualizing these anisotropy terms, providing unique opportunities to study in-depth how molecular geometry (i.e. symmetry) influences magnetic anisotropy. We employ theoretical irreducible tensor operator methods that improve significantly on previous numerical methods applied to trigonal Mn3 clusters in order to gain microscopic insights into the molecular anisotropy. We find that the easy-axis tilting leads to a dramatic compression of the effective energy barrier to magnetization reversal, thus accounting for the absence of single-molecule magnet behavior found in related $Mn_3^{III}$ systems.

## VII. ACKNOWLEDGEMENTS

We thank Nick Bonesteel, Johannes McKay and Frederic Mentink-Vigier for useful suggestions. We acknowledge financial support from the US National Science Foundation (DMR-1610226), the Air Force Office of Scientific Research (Asian Office of Research and Development, contract FA2386-17-1-4040) and the Ministry of Science and Technology of Taiwan (MOST105-2113-M-030-006). A portion of this work was performed at the US National High Magnetic Field Laboratory, which is supported by the National Science Foundation (DMR-1157490 and DMR-1644779) and the State of Florida.

## APPENDIX: IRREDUCIBLE TENSOR OPERATOR METHODS

In this Appendix, we derive the expressions given in Eqs. (8) and (9) using irreducible tensor operator (ITO) methods. Before going further, it is important to introduce some conventions for using ITOs as they apply to the spin Hamiltonian. Though the notation may appear convoluted at first sight, this transformation greatly simplifies the otherwise cumbersome algebraic steps necessary to compute equivalent operators.

A general cartesian operator $\hat{O}(S)$, which is a function of spin operators $S$, can be expanded into ITOs as [34]:

$$\hat{O}(S) = \sum_{k} \sum_{q=-k\ldots+k} c_{kq} \hat{T}_q^{(k)}(S), \qquad (A.1)$$

Here, $k$ denotes the rank of the spherical tensor, with its $q^{\text{th}}$ component running from $-k$ to $k$. Note that the more commonly used Stevens operators[30] representing the higher order ZFS terms in Eq. (2) have the indices denoting rank $k$ and component $q$ swapped, i.e., they are written $B_k^q \hat{O}_k^q$. For compactness and ease of comparison with prior works, the proceeding derivations follow the convention used in [34] and [37] where a general tensor operator's rank '$k$' is in the superscript, while its degree '$q$' is in the subscript, i.e., $\hat{T}_q^k$.

When considering $\hat{H}_1$ in the expansion of the spin Hamiltonian of Eq. (7), we can rewrite one individual $\overleftrightarrow{d}$ tensor product in the most general expansion:

$$\hat{S}_i \cdot d_{ij} \cdot \hat{S}_j = \sum_{k=0,1,2} \sum_{q=-k\ldots+k} \hat{T}_q^{(k)*}(\overleftrightarrow{d}) \hat{T}_q^{(k)}(S) \qquad (A.2)$$

with

$$c_{kq} = (-1)^q \hat{T}^{(k)}_{-q}(\overleftrightarrow{d}) = \hat{T}^{(k)*}_q(\overleftrightarrow{d}) \tag{A.3}$$

However, since $\overleftrightarrow{d}$ is given by a traceless $3 \times 3$ symmetric tensor, we can greatly simplify the sum to only include ITOs of rank $k = 2$. This leaves only three terms in the summation originating from the non-zero contributions to the local $\overleftrightarrow{d}$ anisotropy:[41]

$$T^{(2)}_0(\overleftrightarrow{d}) = \frac{1}{\sqrt{6}}[3d_{zz} - (d_{xx} + d_{yy} + d_{zz})] = \sqrt{\frac{2}{3}}d$$

and,

$$T^{(2)}_{\pm 2}(\overleftrightarrow{d}) = \frac{1}{2}[d_{xx} - d_{yy} \pm i(d_{xy} + d_{yx})] = e \tag{A.4}$$

In this representation, each spin site can be related to its neighbors by specifying Wigner Rotations in place of the Euler rotations, as required by the more familiar Cartesian representation. This is given by an expansion in terms of a linear combination of Wigner matrix elements (see appendix B in [41]):

$$T^{(k)}_q(\overleftrightarrow{d}) = \sum_{q'=-k\ldots+k} \hat{D}^{(k)}_{q'q}(\alpha,\beta,\gamma) T^{(k)}_{q'}(\overleftrightarrow{d}) \tag{A.5}$$

from which, $\hat{H}_1$ can be rewritten as:

$$\hat{H}_1 = \sum_k \sum_{q=-k\ldots k} T^{(k)*}_q(\overleftrightarrow{d})\hat{T}^{(k)}_q(s_1) + T^{(k)*}_q(\overleftrightarrow{d})\hat{T}^{(k)}_q(s_2) + T^{(k)*}_q(\overleftrightarrow{d})\hat{T}^{(k)}_q(s_3) \tag{A.6}$$

where the indices $s_{i=1,2,3}$ have been used to denote Wigner rotations for spin sites where $\alpha = 0°, 120°, 240°$, to preserve the $C_3$ symmetry of the molecule. As an example, if we focus on one Mn site for an arbitrary tilt angle $\beta$, using the '$zyz$' Euler rotation convention,[30] we find for $\alpha = 0, \gamma = 0, k = 2$ and $q = 0$:

$$T^{(2)}_{q=0}(\overleftrightarrow{d}) = \hat{D}^{(2)}_{-2,0}(0,\beta,0) T^{(2)}_{-2}(\overleftrightarrow{d}) + \hat{D}^{(2)}_{0,0}(0,\beta,0) T^{(2)}_0(\overleftrightarrow{d}) + \hat{D}^{(2)}_{2,0}(0,\beta,0) T^{(2)}_2(\overleftrightarrow{d}) \tag{A.7}$$

$$= \frac{d}{\sqrt{6}}(3\cos^2\beta - 1) + e\sqrt{\frac{3}{2}}\sin^2\beta$$

We conveniently obtain this same expression for the Mn sites with $\alpha = 120°$ and $\alpha = 240°$ since $\widehat{D}^{(k=2)}_{q'q=0}(\alpha, \beta, \gamma)$ is independent of $\alpha$.

Concerning projections to first order in perturbation, as briefly introduced in Eq. (5), the replacement theorem can be applied using the Wigner-Eckart formalism to re-express the ITOs as:

$$P_{\tau S}\hat{T}^{(2)}_q(s)P_{\tau S} = \Gamma_2(s)\hat{T}^{(2)}_q(S) \tag{A.8}$$

where the projection coefficient is given by a ratio of reduced matrix elements (given by Wigner 3j symbols, and tabulated in the appendix of Waldmann and Güdel [34]), in which:

$$\Gamma_2(s) = \frac{\langle \tau S||\hat{T}^{(2)}(s)||\tau S\rangle}{\langle S||\hat{T}^{(2)}||S\rangle} \tag{A.9}$$

This conveniently allows us to relate the effective GSA operators in the $|SM\rangle$ basis to the local MS operators in the $|\tau SM\rangle$ basis that spans the full spin-dependent Hilbert space of the cluster. The expression for the second-order projection is a bit more cumbersome, and will only be restated here:

$$P_{\tau S}\hat{T}^{(k_1)}_{q_1}(s_1)P_{\tau'S'}\hat{T}^{(k_2)}_{q_2}(s_2)P_{\tau S} = \sum_k \Gamma_{k_1 k}(s_1)\Gamma^*_{k_2 k}(s_2)d^{k_1 k_2 k}_{q_1 q_2 q}\hat{T}^{(k)}_q(S) \tag{A.10}$$

where we again choose $k_1 = k_2 = 2$ for the MS operators ($\hat{s}$), with $k$ restricted by $q = q_1 + q_2$. The $d^{k_1 k_2 k}_{q_1 q_2 q}$ factors consist of products of 3j and 6j symbols that are once again tabulated in Waldmann and Güdel[34] and do not need to be addressed here. The more general second-order projection coefficient is then given by:

$$\Gamma_{k_1,k}(s) = \frac{\langle \tau S||\hat{T}^{(k_1=2)}(s)||\tau''S''\rangle}{\langle S||\hat{T}^{(k)}||S\rangle} \tag{A.11}$$

which relates the multi-spin '$\hat{s}$' operators of order $k_1 = 2$ to an effective giant spin '$\hat{S}$' operator of order $k$.

**Perturbation with ITOs**

From the above definitions, the perturbation to isotropic exchange can now be rewritten, somewhat compactly, in terms of ITOs:[34]

$$\hat{H}_{\text{eff}} = \sum_q (-1)^q h^{1(k)}_{-q} \hat{T}^{(k)}_q(S) - \sum_{S''} \sum_{k,q} \frac{(-1)^q h^{2(k)}_{-q} \hat{T}^{(k)}_q(S)}{\Delta} \qquad (A.12)$$

In the proceeding sections we will explicitly expand the 1$^{st}$ and 2$^{nd}$ order perturbative terms in order to respectively obtain expressions for the 2$^{nd}$ and 4$^{th}$ order molecular (GSA) anisotropies.

**First Order in Perturbation**

The first order factor $h^{1(k)}_q$ in Eq. (A12) is given by the scalar product of the local $\overset{\leftrightarrow}{d}$ tensors and the 1$^{st}$ order projection coefficients

$$h^{1(k_r)}_{q_r} = \sum_{s_i} \Gamma_{k_r}(s_i) T^{(k_r)}_{q_r}(\overset{\leftrightarrow}{d}) \qquad (A.13)$$

Since most studies of SMMs are generally concerned with molecular ZFS of the form $D\hat{S}^2_z$, we will focus on $T^{(2)}_{q=0}(\overset{\leftrightarrow}{d})$, which will give an expression that can specifically relate the local anisotropy to the 2$^{nd}$ order molecular GSA anisotropy. For a single site, this was derived in Eq. (A12). Then, for $\hat{D}^{(2)}_{q',0}(0,\beta,0) = \hat{D}^{(2)}_{q',0}(120°,\beta,0) = \hat{D}^{(2)}_{q',0}(240°,\beta,0)$, as a function of tilt angle $\beta$, we find:

$$h^{1(k=2)}_{q=0} = D_{mol}(\beta) = 3\Gamma_{k=2}(s) \left[ \frac{d}{2}(3\cos^2\beta - 1) + \frac{3}{2} e \sin^2\beta \right] \qquad (A.14)$$

where the projection factor described in Eq. (A9) has $\Gamma_{k=2}(s) \approx .0909$ for $s = 2$ and $S = 6$ (see expressions in Table 11.9 of Boča [33]).

**S-mixing in Second Order perturbation**

The second order factor $h_q^{2(k)}$ in Eq. (A12) is given by the product of local $\overleftrightarrow{d}$ tensors [34]:

$$h_q^{2(k)} = \sum_{\tau''} \sum_{r,u} \sum_{s_r,s_u} \Gamma_{k_r k}(s_r) \Gamma_{k_u k}^*(s_u) \frac{(-1)^{k_r - k_u} C_{k_r k_u k}}{\sqrt{2k+1}} \times \left[ T^{(k_r)}(V_{s_r}) \otimes T^{(k_u)}(V_{s_u}) \right]_q^{(k)} \quad (A.15)$$

where the tensor product of two general ITOs is written as a sum over Clebsch-Gordan coefficients with the direct product of the individual tensor components:

$$T_Q^{(K)} = \left[ U^{(k_1)} \otimes V^{(k_2)} \right]_Q^{(K)} = \sum_{q_1 q_2} \langle k_1 k_2 q_1 q_2 | KQ \rangle U_{q_1}^{(k_1)} V_{q_2}^{(k_2)} \quad (A.16)$$

in which the uppercase indices pick out the $K^{\text{th}}$ and $Q^{\text{th}}$ components of the total $U$, $V$ product. This permits the computation of each ESO $B_q^k$ coefficient, where the selection rules from the Clebsch-Gordan coefficients determine which local tensor components contribute to the product, i.e., $q_1 + q_2 = Q$.

As applied to Mn$_3$, we now generate an expression for the coefficient of $\hat{O}_4^3$ $\left( = [\hat{S}_z, \hat{S}_+^3 + \hat{S}_-^3]_+ \right)$ from the local $d$ and $e$ anisotropies in order to replicate the 3-fold behavior seen in Fig. 8(b). First working out the tensor product for $K = 4$ and $Q = 3$, the only non-zero terms contributing to the sum have $q_1 + q_2 = 3$. In terms of 3j symbols, for a single site, this yields:

$$\begin{aligned} T_3^{(4)} &= \left[ T^{(2)}(\overleftrightarrow{d}) \otimes T'^{(2)}(\overleftrightarrow{d}) \right]_3^4 \\ &= 3(-1)^3 \left[ \begin{pmatrix} 2 & 2 & 4 \\ 1 & 2 & -3 \end{pmatrix} T_1^{(2)}(\overleftrightarrow{d}) T_2'^{(2)}(\overleftrightarrow{d}) + \begin{pmatrix} 2 & 2 & 4 \\ 2 & 1 & -3 \end{pmatrix} T_2^{(2)}(\overleftrightarrow{d}) T_1'^{(2)}(\overleftrightarrow{d}) \right] \end{aligned} \quad (A.17)$$

A subtle yet extremely important step requires that the tensor product be taken *after* applying the proper Wigner rotations. These rotations, which generate off-diagonal elements in the local $\overleftrightarrow{d}$ tensors, specifically give rise to the trigonal terms in the GSA Hamiltonian. For example, in the above expression, $T_1^{(2)}(\overleftrightarrow{d})$ only becomes nonzero after a rotation about the Euler angle $\beta$. The emergence of this off-diagonal term, which is required to produce a non-zero tensor product for $T_3^{(4)}$, demonstrates a clear symmetry lowering of the Hamiltonian by introducing a $\overleftrightarrow{d}$ tensor tilt of $\beta$, and is solely responsible for producing the trigonal anisotropy terms in the GSA.

Now, since $\hat{D}^{(2)}_{q',2}(0,\beta,0) \neq \hat{D}^{(2)}_{q',2}(120°,\beta,0) \neq \hat{D}^{(2)}_{q',2}(240°,\beta,0)$, the calculation becomes significantly more time consuming. Fortunately, symbolic computation programs such as Mathematica[42] make solving expressions like these much more convenient. When considering the sum over each spin site, the second order factor in Eq. (A15) becomes:

$$\begin{aligned} h_{q=3}^{2(k=4)} &= \sum_{\tau''} \frac{c_{2,2,4}}{3} \Big[ \Gamma_{2,4}(s_1)\Gamma^*_{2,4}(s_1)T_3^{(4)}(Mn_1) + \Gamma_{2,4}(s_2)\Gamma^*_{2,4}(s_2)T_3^{(4)}(Mn_2) \\ &\quad + \Gamma_{2,4}(s_3)\Gamma^*_{2,4}(s_3)T_3^{(4)}(Mn_3) \Big] \\ &= -2\sum_{\tau''} \Gamma_{2,4}(s)\Gamma^*_{2,4}(s)c_{2,2,4}\begin{pmatrix} 2 & 2 & 4 \\ 1 & 2 & -3 \end{pmatrix}\left[\frac{3}{4}(2d^2+e^2)\cos\beta\sin^3\beta\right] \end{aligned} \quad (A.18)$$

where we have simplified the expression by taking advantage of the fact that, in the present symmetry, the contributions to the sum over degenerate states within a given multiplet are equal, i.e., $\sum_{\tau''}\Gamma_{2,4}(s_1)\Gamma^*_{2,4}(s_1) = \sum_{\tau''}\Gamma_{2,4}(s_2)\Gamma^*_{2,4}(s_2) = \sum_{\tau''}\Gamma_{2,4}(s_3)\Gamma^*_{2,4}(s_3) \equiv \sum_{\tau''}\Gamma_{2,4}(s)\Gamma^*_{2,4}(s)$. We can then directly relate this to the 4th order trigonal ESO pre-factor via Eq. (A12) as:

$$B_4^3(\beta) = \sum_{\tau''s''} \frac{3}{2\Delta}\Gamma_{2,4}(s)\Gamma^*_{2,4}(s)d^{2,2,4}_{1,2,3}(2d^2+e^2)\cos\beta\sin^3\beta \quad (A.19)$$

We find that this final expression is slightly different from the one derived for Fe$_3$Cr,[37] for which a simplifying assumption was made to drop the dependence on the Euler angle α from $T_1^{(2)}(\overleftrightarrow{d})$ and $T_2^{(2)}(\overleftrightarrow{d})$ in the molecular frame. However, the two expressions are equivalent in the axial limit where the local $e = 0$.